\documentclass[twocolumn,superscriptaddress]{revtex4-1}
\usepackage{style_for_arxiv}
\usepackage{color}
\usepackage{newcommands}
\newcommand{\citen}[1]{\cite{#1}} 
\begin{document}
\title{Neural Network Approach to Construction of Classical Integrable Systems}
\author{Fumihiro Ishikawa}
\affiliation{Department of Physics, The University of Tokyo, Tokyo 113-0033, Japan}
\author{Hidemaro Suwa}
\affiliation{Department of Physics, The University of Tokyo, Tokyo 113-0033, Japan}
\author{Synge Todo}
\affiliation{Department of Physics, The University of Tokyo, Tokyo 113-0033, Japan}
\affiliation{Institute for Solid State Physics, The University of Tokyo, Kashiwa 277-8581, Japan}
\date{\today}
\begin{abstract}
  Integrable systems have provided various insights into physical phenomena and mathematics.
  The way of constructing many-body integrable systems is limited to few ansatzes for the Lax pair, except for highly inventive findings of conserved quantities.
  Machine learning techniques have recently been applied to broad physics fields and proven powerful for building non-trivial transformations and potential functions.
  We here propose a machine learning approach to a systematic construction of classical integrable systems.
  Given the Hamiltonian or samples in latent space, our neural network simultaneously learns the corresponding natural Hamiltonian in real space and the canonical transformation between the latent space and the real space variables.
  We also propose a loss function for building integrable systems and demonstrate successful unsupervised learning for the Toda lattice.
  Our approach enables exploring new integrable systems without any prior knowledge about the canonical transformation or any ansatz for the Lax pair.
\end{abstract}

\maketitle

Integrable systems have provided deep understandings of physical phenomena and profound insight into the connection between physics and mathematics.
For example, the Toda lattice revealed the solution of the solitary wave called the soliton~\cite{Toda1967,Toda1967-2}, and Calogero-Moser systems---a group of integrable systems constructed by an ansatz for the Lax pair---unveiled the relation between integrability and motions in the Lie group~\cite{Calogero1975,Calogero1976,Moser1975}.
It is essential to find or build a new integrable model to advance the understanding of many-body systems further.
However, it is quite challenging as well: the integrability of correlated systems is highly non-trivial.
So far, the path to finding an integrable system is limited to the ingenious construction of constant motions (conserved quantities) and few ansatzes for the Lax pair.
Thus, it is desirable to establish a systematic way of finding integrable models.

In the meantime, machine learning techniques have been successfully applied to many physical systems~\cite{Carleo-etal2019}.
In particular, the neural network has proven powerful for learning complex transformations~\cite{Noe-etal2019,Shuo-Hui-etal2020,Bondesan-Lamacraft2019} and potential functions~\cite{Greydanus-Dzamba-Yosinki2019,Suwa2019}.
As a relevant application of unsupervised learning, the underlying system Hamiltonian can be rebuilt from the trajectory of particles~\cite{Greydanus-Dzamba-Yosinki2019}.
One of the most promising applications of the neural network is the canonical transformation of classical mechanical systems.
The complex distribution of correlated particles can be transformed into the Gaussian distribution of independent harmonic oscillators, enabling efficient sampling of equilibrium states~\cite{Noe-etal2019,Shuo-Hui-etal2020}.
In the classical integrable system, an extensive number of conserved quantities, namely action variables, exist.
The positions and momenta of the particles in real space are canonically transformed into the action-angle variables $(I,\theta)$ in latent space; accordingly, the Hamiltonian $H(p,q)$ becomes $K(I)$ in latent space through the transformation.
The latent-space Hamiltonian depends only on $I$, and thus the action variables are conserved obviously.
The neural network can reproduce the canonical transformation of some known integrable models~\cite{Bondesan-Lamacraft2019}.
Nevertheless, the application of machine learning to the canonical transformation has been limited to known integrable systems.

In this Letter, we propose a systematic way of constructing integrable systems without any prior knowledge about the canonical transformation or any ansatz for the Lax pair.
Training data in our approach are samples of the action-angle variables.
The action variables are sampled from the Boltzmann distribution, and the angle variables are sampled from the uniform distribution.
We assume the functional form of $K(I)$ or that samples of action-angle variables can be generated in some way.
Under this condition, we train neural networks in unsupervised learning such that the networks learn the canonical transformation and the Hamiltonian $H(p,q)$ simultaneously.
Our neural network seeks the natural Hamiltonian corresponding to the assumed $K(I)$, which consists of the kinetic term and the potential term.
While the potential function is represented using the residual neural network, the canonical transformation is implemented composing the RealNVP neural network, the symplectic linear transformation, and the discrete Hartley transformation.
We also use the symplectic integrator with the adjoint method for time evolution in the learning.
We here demonstrate successful learning for the Toda lattice.
The trained neural networks reproduce the true potential function and the exact canonical transformation with high accuracy.
The present approach enables us to find new integrable systems.

Among several definitions of integrability, we adopt the {\em Liouville-Arnold integrability}~\cite{Arnold-text,Perelomov-text,Arutyunov-text2019}.
Let $N$ be the number of particles and suppose that they move in one-dimensional space.
The dimension of the phase space is $2N$.
We here introduce involution: $N$ smooth functions of a Hamiltonian system denoted by $\{F_i\}_{i=1}^N$ are {\em in involution} if they satisfy the following condition:
\begin{align}
\{F_i, F_j\} = 0 \qquad i,j = 1, \ldots, N,
\end{align}
where $\{\cdot\}$ is the Poisson bracket:
\begin{align}
 \{A, B\} \defeq \sum_{i=1}^N
                 \left[\frac{\partial A}{\partial p_i}\frac{\partial B}{\partial q_i}
                 - \frac{\partial A}{\partial q_i}\frac{\partial B}{\partial p_i}\right].
\end{align}
Let us consider the $N$ smooth functions to be conserved quantities.
Obviously, the Hamiltonian can be one of the functions.
Then, the integrability is defined in the following way: {\em if a Hamiltonian system has $N$ conserved quantities in involution that are independent at every point of the phase space, the system is integrable}.
Here, the independence of the functions means that $dF_i$ are linearly independent, \ie the rank of the Jacobian of $F_i$ is $N$.

Our analysis is based on the property of the integrable system that the trajectory is diffeomorphic to the $N$-dimensional torus if the motion of the particles is bounded or periodic.
Let $\{\theta_i\}_{i=1}^{N}$ be angular coordinates of each dimension of the torus.
The motion of the integrable system is represented by
\begin{align}
   \frac{d\theta_i}{dt} & = \omega_i(F) \qquad i=1,\ldots,N,
\end{align}
where $\{\omega_i(F)\}_{i=1}^{N}$ are functions of $F=\{F_i\}_{i=1}^{N}$.
It is known that a set of variables $\{I_i\}_{i=1}^{N}$ can be constructed from $\{F_i\}_{i=1}^{N}$ such that $\{I_i,\theta_i\}_{i=1}^N$ are canonical coordinates~\cite{Arnold-text}: $I_i$ and $\theta_i$ are called the action and angle variables, respectively.
The equations of motion are given by
\begin{align}
 \begin{split}
   \frac{dI_i}{dt} & = 0, \\
   \frac{d\theta_i}{dt} & = \frac{\partial K(I)}{\partial I_i} = \omega_i(I), \qquad i=1,\ldots,N,
 \end{split} \label{eq:eom_of_action_angle_variabels}
\end{align}
where $K(I)$ is the Hamiltonian in latent space, which depends only on the action variables.
See Supplemental Material for some examples of the action-angle variables~\cite{our_supplemental_material}.

Equation~(\ref{eq:eom_of_action_angle_variabels}) implies that the system is integrable if the corresponding action-angle variables exist~\cite{Arnold-text}.
Is there any reasonable condition on the Hamiltonian? We seek systems that can be represented by a natural Hamiltonian~\cite{Cariglia-2014} in real space, which consists of the kinetic term and a potential function, that is,
\begin{align}
  H(p,q) = \sum_{i=1}^{N}\frac{p_i^2}{2} + V(q),\label{eq:natural_hamiltonian}
\end{align}
where $p=\{p_i\}_{i=1}^{N}$ are momenta, $q=\{q_i\}_{i=1}^{N}$ are positions or displacements, and $V(q)$ is a potential function.
An integrable system described by $K(I)$ is {\em physically reasonable} if the Hamiltonian can be represented by the natural form~(\ref{eq:natural_hamiltonian}) in real space.
In general, however, it is non-trivial to find an appropriate canonical transformation from a given $K(I)$ to a natural $H(p,q)$.
Thus, we propose a neural network approach to construction of the canonical transformation and the natural Hamiltonian simultaneously.

\begin{figure}[tp]
  \centering
  \includegraphics[width=8.5cm]{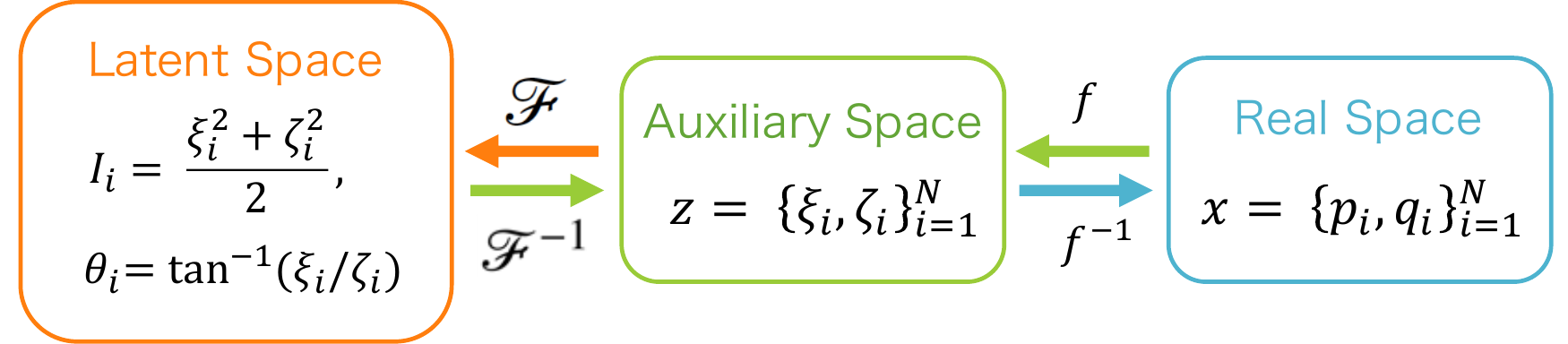}
  \caption{
    (Color online) Real-space, auxiliary-space, latent-space variables, and the canonical transformations between them.
    }
\label{fig:map_of_transformations}
\end{figure}
The classical integrable system has $N-1$ action variables and the total momentum as conserved quantities.
We define $(I_N, \theta_N)$ as
\begin{align}
  \begin{split}
    I_N & = \frac{1}{2N}\left(\sum_{i=1}^N p_i\right)^2, \\
    \theta_N & = \frac{1}{\sqrt{N}}\sum_{i=1}^N q_i,
  \end{split}
\end{align}
and assume the total momentum to be zero to avoid a drift of the center of mass.
We also assume the potential function to be translation invariant and represented by the following form:
\begin{align}
    V(q) & = \sum_{i=1}^{N}v(r_{ii+1}), \label{eq:chain}
\end{align}
where $r_{ij}$ is the difference between the displacements of the adjacent particles $r_{ij} = q_i - q_j$.

For convenience, we call the space described by the canonical coordinates $x = \{p_i,q_i\}_{i=1}^{N}$ the real space.
We also introduce the auxiliary space variables $z = \{\xi_i,\zeta_i\}_{i=1}^{N}$, which are related to the action-angle variables by
\begin{align}
  \begin{split}
    I_i & = \frac{\xi_i^2 + \zeta_i^2}{2}, \\
    \theta_i & = \tan^{-1}\left(\frac{\xi_i}{\zeta_i}\right), \qquad i = 1, \ldots N-1,
  \end{split} \label{eq:latent_space_representation_of_action_angle_variables}
\end{align}
and $(\xi_N,\zeta_N)$ are the momentum and the position of the center of mass, respectively.
We introduce the auxiliary variables to represent the bounded property of the system and the conservation of the torus radii straightforwardly.
Let $\mathscr{F}$ and $f$ be the canonical transformations between the real-space, the auxiliary-space, and the latent-space variables, as shown in \reffig{fig:map_of_transformations}. We represent $f$ using a neural network~\cite{our_supplemental_material}.
The composite transformations, $\mathscr{F} \circ f$ and $f^{-1}\circ \mathscr{F}^{-1}$, are also canonical transformations.
The canonical transformation from the real-space variables into the action-angle variables was introduced in Ref.~\citen{Bondesan-Lamacraft2019}.
The existence of the action-angle variables was proven in some integrable systems~\cite{Nguyen2005,Ito1989,Kappeler1998,Kappelerf-Henricit2008,Henrici2015}.
\begin{figure}[tp]
  \centering
  \includegraphics[width=8cm]{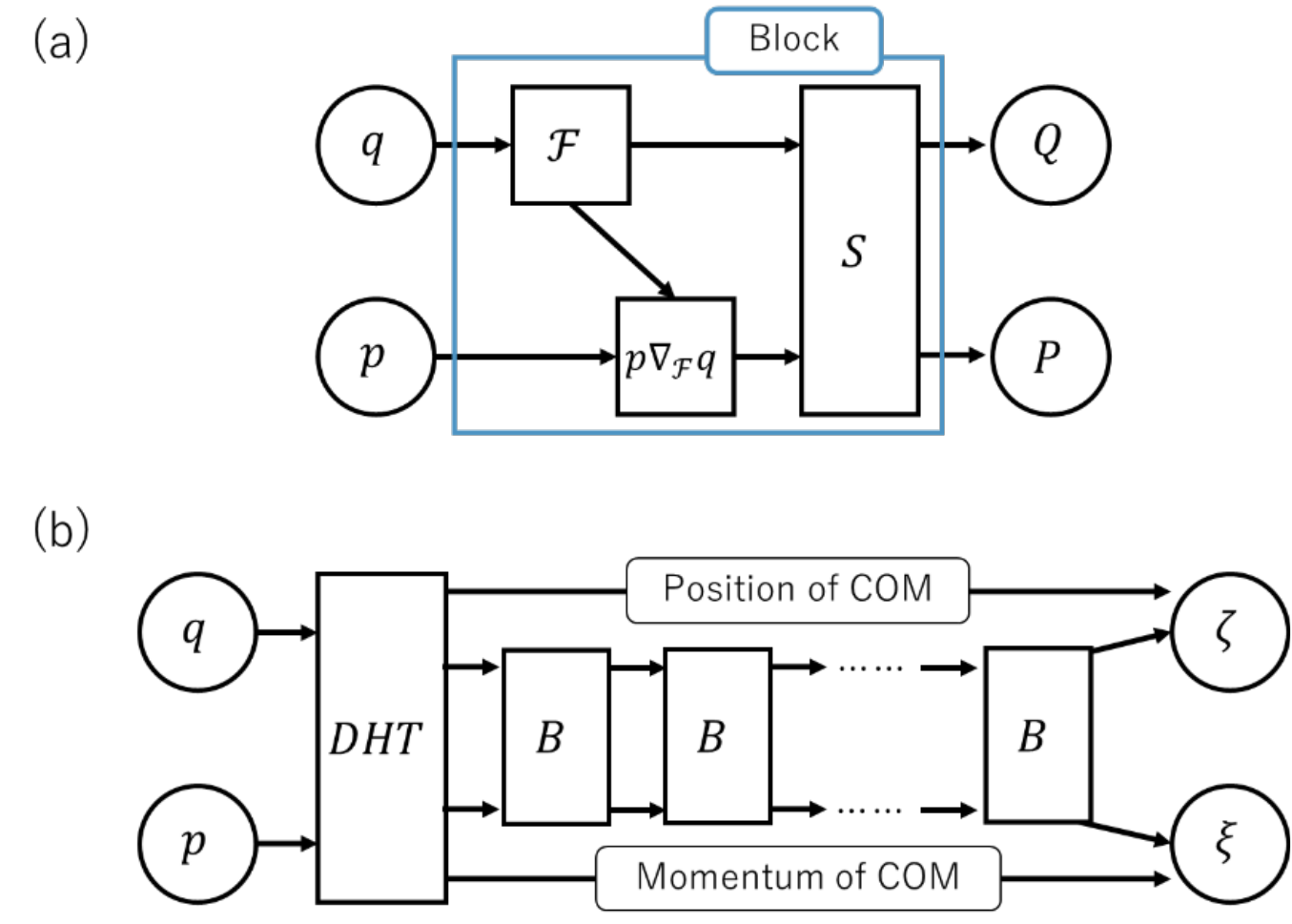}
  \caption{
    Schematic picture of our canonical transformations from real space to auxiliary space.
    (a) An elemental block constructing our canonical transformation.
    This block is a combination of a point transformation and a symplectic linear transformation.
    $\mathcal{F}$ is a bijective map represented by RealNVP.
    The transformation of $p$ is derived from the generator function obtained from the point transformation and
    $S$ stands for the symplectic linear transformation.
    (b) The whole structure of our canonical transformation.
    DHT stands for the discrete Hartley transformation and
    $B$ means the elemental block defined in (a).
    COM stands for the center of mass.
    }
  \label{fig:canonical_trans_diagrams}
\end{figure}

We represent $v(r)$ in \refeq{eq:chain} by using the residual neural network~\cite{He-etal2016,Sehanobish-etal2020}.
The transformation $f$ is composed of three transformations: the canonical transformation generated by point transformations represented by the neural network~\cite{Shuo-Hui-etal2020}, the symplectic linear transformation parameterized by the Iwasawa decomposition~\cite{Bondesan-Lamacraft2019,Iwasawa1949}, and the discrete Hartley transformation~\cite{Hartley1942,Bracewell1983}.
A schematic picture of our canonical transformations is illustrated in \reffig{fig:canonical_trans_diagrams}.
The point transformations are implemented using the RealNVP neural network, which is one of the invertible neural networks~\cite{RealNVP2017}.
The invertible neural network has the universal approximation property under some conditions~\cite{Teshima-et-al2020}.
Since the point transformation acts only on position coordinates, we introduce a
symplectic transformation to represent the coupling between position
and momentum coordinates. We thus expect our neural network to
possess high representability.
The discrete Hartley transformation is used for extracting the motion of the center of mass from the real space coordinates.
See Supplemental Material for details of the neural network and parameter settings~\cite{our_supplemental_material}.
We used the Adam optimizer~\cite{Adam2017}.
The learning rate was reduced after some epochs to improve the accuracy.
For the other hyperparameters in Adam, we used the same values proposed in Ref.~\citen{Adam2017}.

Training data in our approach are composed of samples of the action-angle variables $\{I_i,\theta_i\}_{i=1}^{N}$.
The action variables are sampled from the Boltzmann distribution $\rho_{\mathrm{B}}$,
\begin{align}
  \rho_{\mathrm{B}} = \frac{e^{-\frac{K(I)}{T}}}{Z}, \; Z = \int_0^{\infty} dI \;e^{-\frac{K(I)}{T}},
\end{align}
where $T$ is the temperature.
The angle variables are sampled from the uniform distribution.

We propose a loss function that consists of two parts: the loss of the action-variable conservation and the loss of the energy equivalence.
Specifically, the losses are given by the mean squared logarithmic error (MSLE) function, which is arguably one of the most useful loss functions for time series analysis~\cite{Liu-etal2017,Zhou-Huang2019,Van-etal2018},
\begin{align}
  \begin{split}
    L & = L_{I} + L_E , \\
    L_I & = \sum_{i=1}^{N-1}\sum_{k=1}^{N_{\mathrm{time}}}
    \frac{\average{|\log(I_i(t_0)+1) - \log(I_i(t_k)+1)|^2}_D}{(N-1)N_{\mathrm{time}}}, \\
    L_E & = \average{|\log(H(p,q)+1) - \log(K(I)+1)|^2}_D,
  \end{split} \label{eq:final_loss_function}
\end{align}
where $N$ is the number of particles, $N_\mathrm{time}$ is the number of time points, and $\average{\cdot}_D$ is the average over the input data.
The loss function $L_E$ quantifies the energy equivalence: the energy is invariant under the canonical transformation.
The loss function $L_I$ computes the difference between the action variables at two-time points.
We calculate the time evolution of the real space variables using the neural network potential function.
The initial state of the time evolution is set through the inverse canonical transformation from training data labeled as $\{I_i(t_0)\}_{i=1}^{N}$.
Then, $\{I_i(t_k)\}_{i=1}^{N}$ are calculated from the time series of $\{p_i,q_i\}_{i=1}^{N}$ through the canonical transformation.
In the present study, we used the adjoint method with a symplectic integrator~\cite{Sanz-Serna2016,NeuralODE} to reduce the memory consumption in the time evolution.

We demonstrate successful learning, applying our approach to the Toda lattice~\cite{Toda1967-2,Toda1967}, which is a prototype of the classical integrable system.
\begin{figure}[tp]
  \centering
  \includegraphics[width=8.4cm]{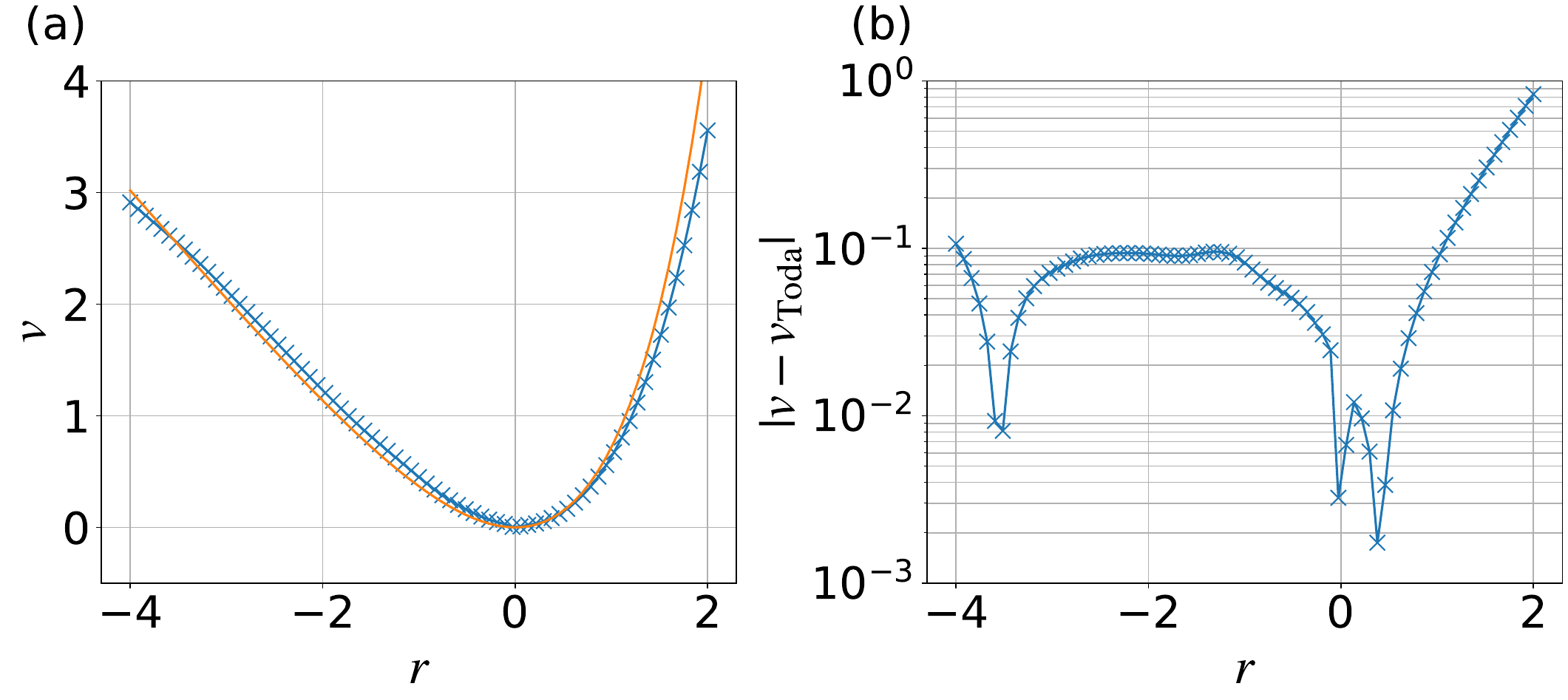}
  \caption{
    (Color online) Neural network potential function obtained from the dataset of action-angle-variable samples at temperature $T=2J$.
    (a) The learned potential function (crosses) and the true Toda potential function (orange solid line) for $J = 1$ and $\alpha = 1$.
    (b) The absolute error of the learned potential function.
    }
\label{fig:toda_larning_residual_result_second_potential}
\end{figure}
\begin{figure}[tp]
  \centering
  \includegraphics[width=8cm]{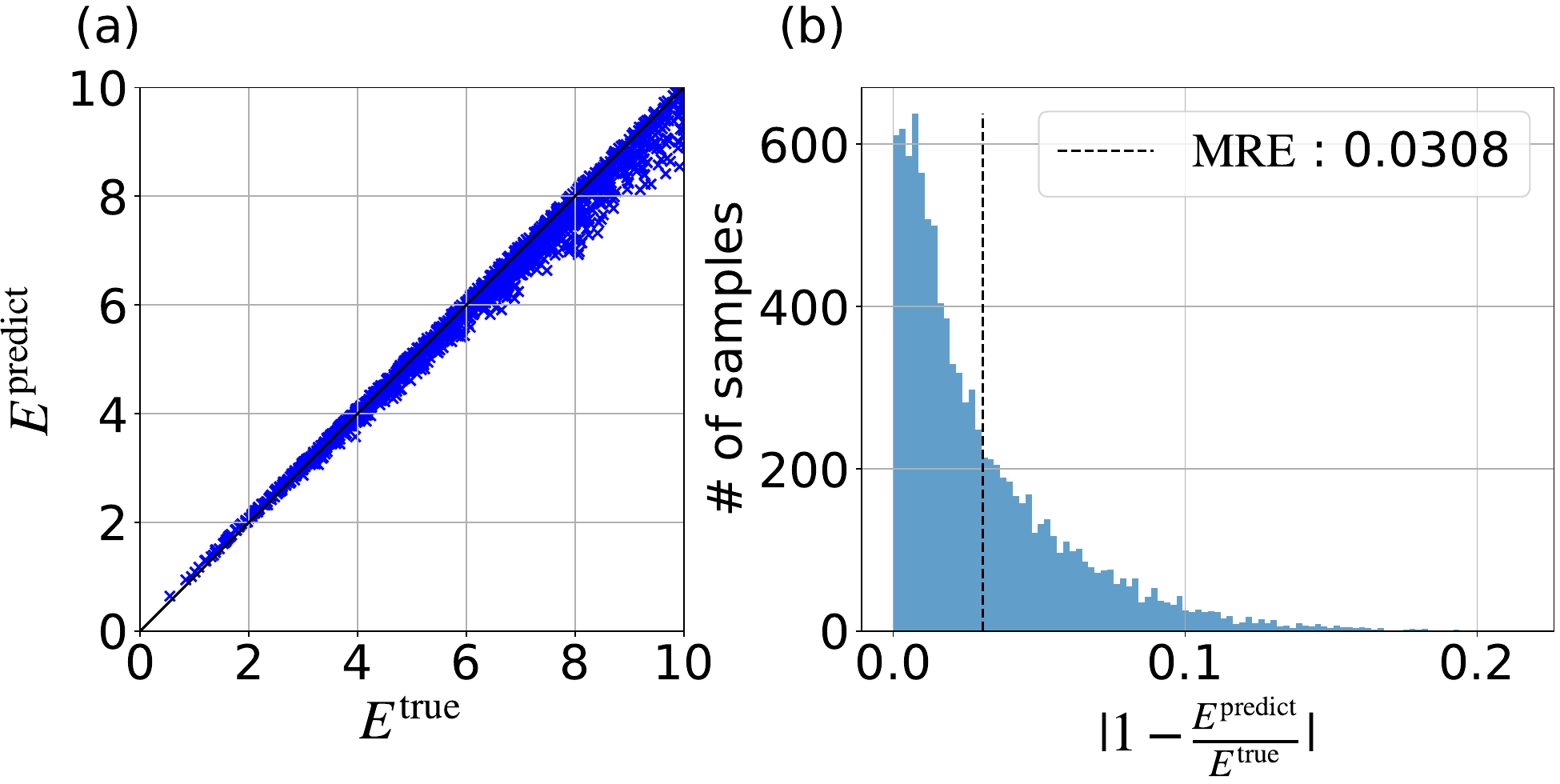}
  \caption{
    (Color online) (a) The scatter plot and (b) the relative error distribution of neural network predictions of the total energy.
    The MRE is 0.0308, averaged over $10^4$ samples.
    }
\label{fig:toda_larning_residual_result_second_energy_error}
\end{figure}
\begin{figure}[tp]
  \centering
  \includegraphics[width=7.7cm]{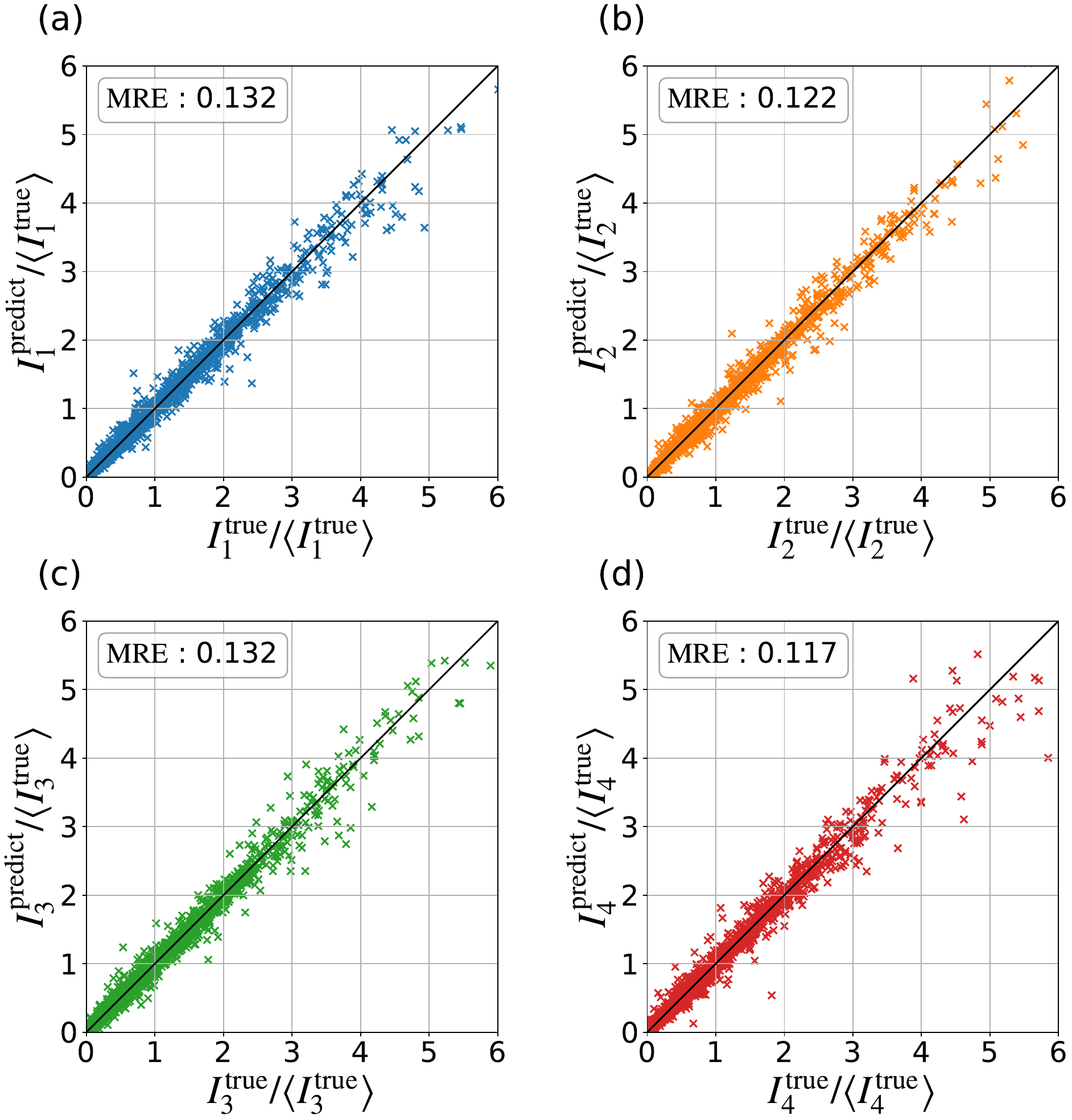}
  \caption{
    (Color online) Scatter plots of neural network predictions of the action variables (a) $I_1$, (b) $I_2$, (c) $I_3$, and (d) $I_4$.
    Using the known analytical expression, we calculated the true values from real space variables sampled from the Boltzmann distribution at $T=2J$.
    The MREs were $0.132$, $0.122$, $0.132$, and $0.117$, respectively, averaged over $10^4$ samples.
    }
\label{fig:toda_larning_residual_result_action_variables_rel_err}
\end{figure}
\begin{figure}[tp]
  \centering
  \includegraphics[width=8cm]{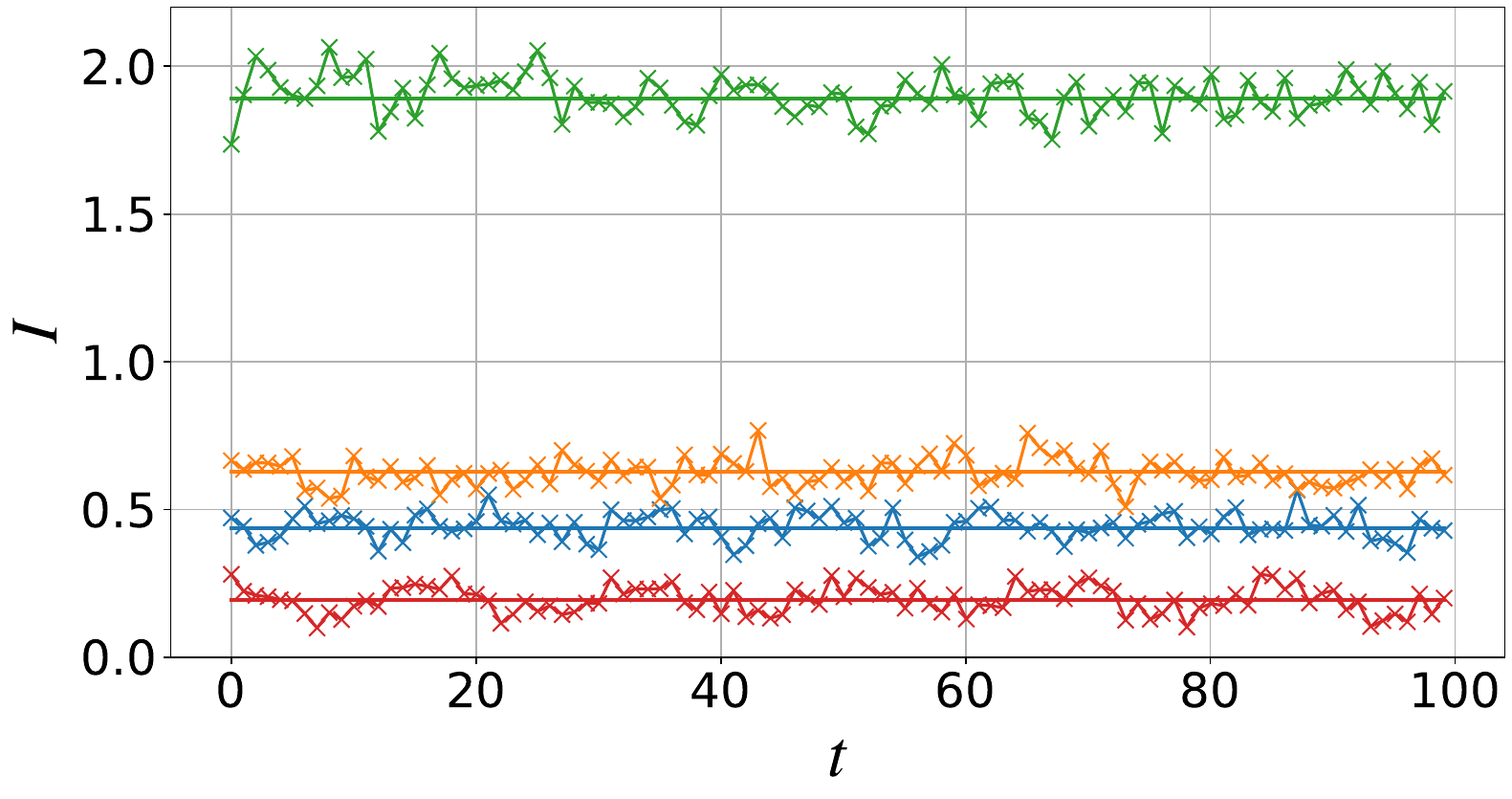}
  \caption{
      (Color online) Time series of action variables predicted by the trained neural networks: $I_1$ (blue), $I_2$ (orange), $I_3$ (green), and $I_4$ (red). The solid lines show the true values.
    }
\label{fig:toda_larning_residual_result_second_action_variables}
\end{figure}
The Hamiltonian is given by
\begin{align}
 H = \sum_{i=1}^{N} \frac{p_i^2}{2m} + \sum_{i=1}^N J\left(e^{-\alpha(q_{i+1} - q_{i})} + \alpha (q_{i+1} - q_{i}) - 1 \right), \label{eq:toda_hamiltonian}
\end{align}
where $J$ and $\alpha$ are the coupling constants, and $q_i$ is the displacement of the $i$-th particle. We set $N=5$, $m=1$, $J = 1$, and $\alpha = 1$ and used the periodic boundary condition.
For the Toda lattice, the function form of $K(I)$ is not known, but the canonical transformation into the action variables is known~\cite{Flaschka-McLaughlin1976,Flaschka1974,Kappelerf-Henricit2008,Henrici2015}.
We first sampled real-space coordinates from the Boltzmann distribution, using the Hamiltonian Monte Carlo (HMC) method for $H(p,q)$~\cite{Duane-etal-1987,mcmc-handbook2011}.
We then obtained samples of the action variables from the real-space coordinates through the known canonical transformation.
Note that we used the exact canonical transformation only for generating input data and testing the final result.
The resulting distribution of the samples is identical to the Boltzmann distribution.
If we had the function form of $K(I)$, we could generate samples from the Boltzmann distribution directly.
The angle variables were sampled from the uniform distribution.
The total momentum, namely $I_5$, was fixed to zero.
The temperature $T$ was set to $2J$.
This temperature is high enough to see the deviation of the Toda lattice potential from the harmonic oscillator and low enough to learn the potential bottom accurately.

We generated $10^5$ samples in total and set the mini-batch size to $2 \times 10^3$~\cite{our_supplemental_material}. We updated the neural network parameters in $160$ epochs.
The learning rate was first set to $10^{-3}$ and reduced to $10^{-4}$ after $80$ epochs.
We calculated the time evolution up to $t = 5$ with $dt = 0.05$ and set $N_\mathrm{time}=5$ at equal intervals for the loss function.

Our neural network successfully reproduced the true potential function, as shown in \reffig{fig:toda_larning_residual_result_second_potential}. It is reasonable that the error becomes larger for higher energy because of the fewer number of samples in the input data.
Note that the learning of the potential function has uncertainty due to periodic boundaries: the total energy is unchanged by adding a linear term as $v'(r) = v(r) + a r$, where $a$ is a constant.
We here plot the potential after removing the linear term: $ \hat{v}(r) - r\left.\frac{\partial \hat{v}(r)}{\partial r}\right|_{r=0}$, where $\hat{v}(r)$ is the learned potential function. See Supplemental Material for details of the neural network parameters~\cite{our_supplemental_material}.
The error of the neural network prediction of the total energy is shown in \reffig{fig:toda_larning_residual_result_second_energy_error}. Here, the mean relative error (MRE) of a quantity $O$ is defined as
\begin{align}
  \mathrm{MRE} = \average{\left|1 - \frac{O^{\mathrm{predict}}}{O^{\mathrm{true}}}\right|}_{D}, \label{eq:mean_relative_error}
\end{align}
where $O^{\mathrm{predict}}$ is the neural network prediction, and $O^{\mathrm{true}}$ is the true value. The average was taken over the dataset.
The MRE of the total energy was only 3\%.

The trained neural network accurately predicted the values of action variables from real space coordinates, as shown in \reffig{fig:toda_larning_residual_result_action_variables_rel_err}.
The true values were calculated from the known analytical expression~\cite{Flaschka-McLaughlin1976,Flaschka1974}.
The MRE of the prediction of the action variables was of order 10\%.
We plot time series of neural network predictions of the action variables in \reffig{fig:toda_larning_residual_result_second_action_variables}.
The time evolution was simulated using the exact potential function in this test.
The predicted action variables do not show any drift with time, fluctuating around the true value.
This result clearly shows that the neural networks learn the conserved quantities without any prior knowledge about the canonical transformation.


In conclusion, we have proposed a machine learning approach to finding classical integrable systems.
Our approach is distinct from the previous approaches in that we construct a natural Hamiltonian from samples in latent space.
We simultaneously train the neural networks to represent the canonical transformation and the potential functions.
In our approach, input data are samples of the action-angle variables.
The action variables are sampled from the Boltzmann distribution, and the angle variables are sampled from the uniform distribution.
We have also proposed a loss function that consists of the two parts: the loss of the action-variable conservation with time and the loss of the energy equivalence.
Note that we simulated the time evolution using the symplectic integrator with the adjoint method.
We applied our approach to the Toda lattice and demonstrated successful unsupervised learning of the canonical transformation and the potential function.
The relative errors of neural network predictions were of order a few percent for the total energy and of order ten percent for the action variables.
Our neural network learned the conserved quantities with no prior knowledge about the canonical transformation.

We here discuss possible extensions of the present approach.
Although we focused on the two-body potential function given by \refeq{eq:chain}, we can extend the function form to more general forms, such as fully connected potential functions.
We can study multi-dimensional systems by applying the discrete Hartley transformation in each dimension.
The entire learning could be improved by adding a loss function to require the distribution of $\{p_i\}_{i=1}^{N}$ to be the Gaussian distributions.
We transformed the real space variables into the latent space variables through the auxiliary space and assumed the specific transformation~(\ref{eq:latent_space_representation_of_action_angle_variables}) between the latent space and the auxiliary space.
Using the auxiliary space is helpful for the learning in the present study.
Nevertheless, if it was removed, the neural network could seek direct canonical transformations between the latent space and the real space.
Whether using the auxiliary space is beneficial may depend on the structure of the neural network.

We also discuss how to distinguish nonintegrability from learning failure.
The loss may remain finite due to the absence of the corresponding integrable model or the low representability of the neural network we use.
It is not easy to distinguish the two cases in practice, which is a typical problem of machine learning approaches. Nevertheless, we can check integrability by monitoring the loss while increasing the neural network representability.
The representability can be enhanced by increasing the number of neural network parameters, such as the network depth and width.
If an integrable natural Hamiltonian of the assumed form exists, the loss should decrease and eventually reach zero as the representability is enhanced.
On the other hand, we can conclude that the given action-variable Hamiltonian is not transformed into the assumed form of natural Hamiltonians if the loss is not decreased by increasing the number of network parameters.

Finally, it is of great interest to search function forms $K(I)$ that allow the Hamiltonian to take the natural form in real space.
It is known that the harmonic oscillator is transformed to a linear function of $K(I)$, and the Hamiltonian with a box (square-well) potential is transformed to a quadratic function of $K(I)$~\cite{Reichl-1992}.
It is, thus, intriguing to find integrable systems described by higher-order functions of $K(I)$ as future problems.

\section*{ACKNOWLEDGEMENTS}
This calculation has been done using NVIDIA GPGPU at Institute for Physics of Intelligence ($i\pi$), the University of Tokyo.
F.I. is supported by the Japan Society for the Promotion of Science through the Program for Leading Graduate Schools (MERIT).

\newpage
\bibliography{bibliography}
\end{document}


\title{Supplemental Material for ``Neural Network Approach to Construction of Classical Integrable Systems''}
\author{Fumihiro Ishikawa}
\affiliation{Department of Physics, The University of Tokyo, Tokyo 113-0033, Japan}
\author{Hidemaro Suwa}
\affiliation{Department of Physics, The University of Tokyo, Tokyo 113-0033, Japan}
\author{Synge Todo}
\affiliation{Department of Physics, The University of Tokyo, Tokyo 113-0033, Japan}
\affiliation{Institute for Solid State Physics, The University of Tokyo, Kashiwa 277-8581, Japan}
\date{\today}
\maketitle

\section{Simple Examples of Action-angle Variables}
The action-angle variables are readily obtained if the Hamiltonian is separable to $N$ decoupled Hamiltonians:
\begin{align}
  H = \sum_{i=1}^{N} H_i(p_i,q_i),
\end{align}
where $H_i$  is the Hamiltonian for each particle.
We consider the case where each motion generated by each partial Hamiltonian $H_i$ is periodic.
In this situation, the action variables are defined by
\begin{align}
  I_i = \frac{1}{2\pi}\oint p_i \, dq_i \qquad{i=1,\ldots,N}, \label{eq:action_variables_circle}
\end{align}
where the contour integration is carried out on the trajectory.
The generating function of canonical transformations of the action variables, denoted as $W$, is given by
\begin{align}
  \begin{split}
    & W(q, I)  = \sum_{i=1}^N W_i(q_i,I_i),
  \end{split}
  \label{eq:generating_function}
\end{align}
where
\begin{align}
  W_i(q_i,I_i) = \int p_i(q_i, I_i) \, dq_i  \qquad{i=1,\ldots,N}.
\end{align}
From the generating function~(\ref{eq:generating_function}), the momenta and the angle variables are obtained as
\begin{align}
  p_i &= \frac{\partial W}{\partial q_i}, \\
  \theta_i &= -\frac{\partial W}{\partial I_i}  \qquad{i=1,\ldots,N},
\end{align}
respectively.

One of the simplest examples is a harmonic oscillator chain of $N$ particles with periodic boundary conditions~\cite{Arnold-text}.
It is well known that the system can be transformed to $N-1$ independent harmonic oscillators and the free motion of the center of mass:
\begin{align}
 H = \frac{1}{2N}\left(\sum_{i=1}^N p_i\right)^2 + \sum_{k=1}^{N-1} \Big[ \frac{P_k^2}{2} + \frac{\omega_k^2 Q_k^2}{2} \Big],
\end{align}
where $\omega_k \defeq \sqrt{4J}\sin (\pi k/N)$ ($k=1,\ldots,N-1$) are the normal mode frequencies with $J$ being the coupling constant, and $\{P_k,Q_k\}_{k=1}^{N-1}$ are the momenta and positions of normal modes.
Here, one can introduce the conserved quantities, which correspond to the ``radii'' of the trajectory in the phase space:
\begin{align}
 \begin{split}
   I_k & = \frac{P_k^2}{2\omega_k} + \frac{\omega_k Q_k^2}{2} \qquad k = 1, \ldots N-1,\\
   I_N & = \frac{1}{2N}\left(\sum_{i=1}^N p_i\right)^2.
 \end{split}\label{eq:harmonic_oscillator_canonical_trans}
\end{align}
The last conserved quantity, $I_N$, always exists when the potential function is translation invariant.
Since the transformation from the original coordinates to the decoupled oscillators is a canonical transformation, it is easy to prove that the quantities are $N$ linearly independent first integrals in involution.
The angular variables are given by
\begin{align}
 \theta_k = \tan^{-1}\left(\frac{P_k}{\omega_k Q_k}\right) \qquad k = 1, \ldots N-1.
\end{align}
The variables $\{I_k,\theta_k\}_{k=1}^{N-1}$ are the canonical coordinates.
Consequently, the harmonic oscillator chain is completely integrable and has the action-angle variables, and $K(I)$ is given by
\begin{align}
  K(I) = \sum_{k=1}^{N-1}\omega_k I_k + I_N \label{eq:harmonic_oscillator_K_I}.
\end{align}
Note that $I_N$ is not the action variable because the corresponding motion is not bounded.
We can also obtain the action variables from the definition, \refeq{eq:action_variables_circle}.
We denote each energy of the periodic motion as
\begin{align}
  E_k = \frac{P_k^2}{2} + \frac{\omega_k^2 Q_k^2}{2} \qquad k = 1, \ldots N-1.
\end{align}
Then, the contour integration~\refeq{eq:action_variables_circle} is evaluated as
\begin{align}
  \begin{split}
    I_k & = \frac{1}{2\pi}\oint P_k \, dQ_k
        = \frac{1}{\pi} \int_{-\frac{\sqrt{2E_k}}{\omega_k}}^{\frac{\sqrt{2E_k}}{\omega_k}} \sqrt{2E_k -\omega_k^2Q_k^2} \, dQ_k
        = \frac{E_k}{\omega_k} \qquad k=1,\ldots,N-1,
  \end{split} \label{eq:action_of_harmonic_derived_from_circle_integration}
\end{align}
which is nothing but the quantities defined by \refeq{eq:harmonic_oscillator_canonical_trans}.

The second example is the square-well potential~\cite{Reichl-1992}.
While $K(I)$ of the harmonic oscillator chain is given by a linear combination of $\{I_i\}_{i=1}^{N-1}$, that for the square-well potential is realized by a linear combination of $\{I_i^2\}_{i=1}^{N-1}$.
Let us introduce the following Hamiltonian:
\begin{align}
  H = \frac{1}{2N}\left(\sum_{i=1}^N p_i\right)^2 + \sum_{k=1}^{N-1} \Big[ \frac{P_k^2}{2} + U(Q_k) \Big], \label{eq:hamiltonian_potential_well}
\end{align}
where $U$ is the square-well potential given by
\begin{align}
  U(r) = \begin{cases}
            0 & -\frac{\pi}{2\sqrt{2}} < r < \frac{\pi}{2\sqrt{2}}, \\
            \infty & \mathrm{otherwise}.
         \end{cases} \label{eq:potential_well}
\end{align}
The system is also separable to $N$ independent motions.
We introduce the energy of each motion as
\begin{align}
  E_k = \frac{P_k^2}{2} + U(Q_k) \qquad k = 1, \ldots N-1.
\end{align}
In the same way as \refeq{eq:action_of_harmonic_derived_from_circle_integration}, the action variables are introduced as
\begin{align}
  \begin{split}
    I_k & = \frac{1}{2\pi}\oint P_k \, dQ_k
        = \frac{1}{\pi} \int_{-\frac{\pi}{2\sqrt{2}}}^{\frac{\pi}{2\sqrt{2}}} \sqrt{2\left(E_k-U(Q_k)\right)} \, dQ_k
        = \sqrt{E_k} \qquad k=1,\ldots,N-1.
  \end{split}
\end{align}
Therefore, $K(I)$ of \refeq{eq:hamiltonian_potential_well} is given by
\begin{align}
 K(I) = \sum_{k=1}^{N-1} I_k^2 + I_N \label{eq:potential_well_K_I}.
\end{align}

For the Toda lattice, the action variables are calculated from the Lax pair~\cite{Flaschka-McLaughlin1976,Flaschka1974}.
We consider the Toda lattice with periodic boundary conditions.
The representative form of the Lax pair of the Toda lattice is~\cite{Flaschka1974-2}
\begin{align}
  L & \defeq
  \left(
    \begin{array}{ccccccccc}
      b_1 & a_1 &       &        &         &         &       &         & a_N     \\
      a_1 & b_2 &       &        &         &         &       &         &         \\
          &     & \ddots &       &         &         & 0     &         &        \\
          &     &       &b_{i-1} & a_{i-1} &         &       &         &         \\
          &     &       &a_{i-1} & b_{i}   & a_i     &       &         &         \\
          &     &       &        & a_{i}   & b_{i+1} &       &         &         \\
          &     &  0    &        &         &         & \ddots &        &        \\
          &     &       &        &         &         &       & b_{N-1} & a_{N-1} \\
      a_N &     &       &        &         &         &       & a_{N-1} & b_{N}
    \end{array}
  \right) \label{eq:toda_lax_matrix}\\
  M & \defeq
  \left(
    \begin{array}{ccccccccc}
      0   &-a_1 &       &        &         &         &       &         & a_N     \\
      a_1 & 0   &       &        &         &         &       &         &         \\
          &     & \ddots &       &         &         & 0     &         &        \\
          &     &       &    0   & -a_{i-1}&         &       &         &         \\
          &     &       &a_{i-1} &   0     & -a_i    &       &         &         \\
          &     &       &        & a_{i}   &   0     &       &         &         \\
          &     &  0    &        &         &         & \ddots&         &        \\
          &     &       &        &         &         &       &    0    & - a_{N-1} \\
    - a_N &     &       &        &         &         &       & a_{N-1} & b_{N}
    \end{array}
  \right),
\end{align}
where $a_n$ and $b_n$ are the Flashka variables defined by~\cite{Flaschka1974}
\begin{align}
  \begin{split}
    a_i & \defeq \frac{1}{2}e^{-(q_{i+1} - q_i)}, \\
    b_i & \defeq \frac{1}{2}p_i \qquad i=1,\ldots,N.
  \end{split}
\end{align}
The matrices, $L$ and $M$, are called the Lax pair and satisfy the following equation~\cite{Lax1968}:
\begin{align}
 \frac{dL}{dt} = [M, L],\label{eq:eom_lax_pair}
\end{align}
where $[\cdot]$ denotes the commutator.
The action variables are calculated by the Lax pair as~\cite{Flaschka-McLaughlin1976,Flaschka1974}
\begin{align}
  I_k = \frac{2}{\pi} \int_{\lambda_{2k}}^{\lambda_{2k+1}} \cosh^{-1}\left[\frac{(-1)^k}{2} \Delta(\lambda)\right] \, d\lambda \qquad k=1,\ldots,N-1, \label{eq:exact_toda_actions}
\end{align}
where $\{\lambda_{2k-1}\}_{k=1}^{N}$ are the eigenvalues of $L$, and $\{\lambda_{2k}\}_{k=1}^{N}$ are those of the matrix obtained from $L$ by replacing $L_{N1}$ and $L_{1N}$ by $-L_{N1}$ and $-L_{1N}$, respectively. We assume that the eigenvalues are sorted in ascending order.
The function $\Delta(\lambda)$ in \refeq{eq:exact_toda_actions} is related to the determinant of $L$ by
\begin{align}
  \mathrm{det}(L - \lambda I) = (-1)^N\left(\prod_{j=1}^{N}a_j\right) \left(\Delta(\lambda) - 2\right), \label{eq:discriminant}
\end{align}
where $I$ is the $N \times N$ identity matrix.
The action variables play the role of the amplitude of the solitons~\cite{Ferguson-Flaschka-McLaughlin1982}.

\section{Structures of Neural Networks}
\begin{figure}[t]
  \centering
  \includegraphics[width=8cm]{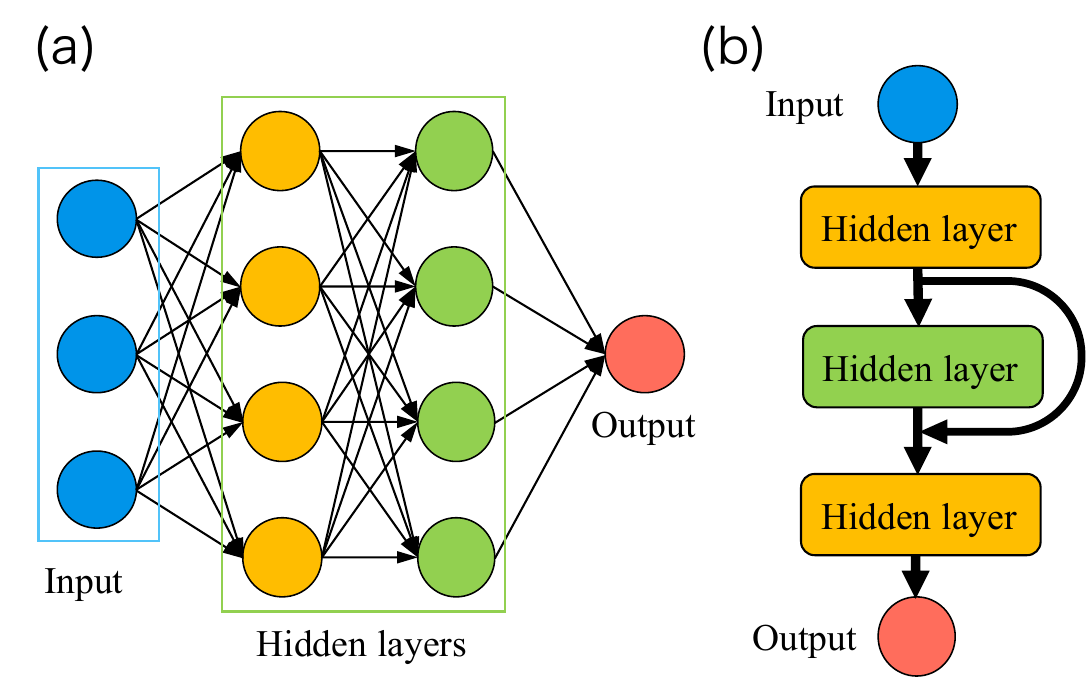}
  \caption{
    (Color online) Schematic diagrams of (a) dense neural networks and (b) residual neural networks.
    }
  \label{fig:residual_and_dense}
\end{figure}
In this section, we describe the details of our neural networks.
We consider a particular group of potential functions, which depends on the difference between two positions $q_i - q_j$.
According to the translational invariance, the motion of the center of mass is trivially conserved.
We assume the following potential function:
\begin{align}
    V(q) & = \sum_{i=1}^{N}v(r_{ii+1}),
\end{align}
where $v(r)$ is represented by a neural network, $N$ is the number of particles, and $r_{ij}$ is the difference between the displacements of the adjacent particles $r_{ij} = q_i - q_j$.
In the present study, $N$ was set to $5$.
We consider two different types of neural networks for $v(r)$, a dense neural network and a residual neural network (see \reffig{fig:residual_and_dense}).
These neural networks have two parameters: the number of hidden layers and the dimension of the hidden layers.
In general, each hidden layer can be different from the others.
In the present work, we used the same dimension for all the hidden layers for simplicity and convenience.
We set the dimension of the hidden layers to $128$ and the number of hidden layers to $4$.
We used the sigmoid-weighted linear units as the activation function~\cite{Elfwing-Uchibe-Doya2018}.
For the Toda lattice, from preliminary tests using different combinations of neural networks, we chose a residual neural network without additive biases.

Our canonical transformation is a map from the real space $\{p_i,q_i\}_{i=1}^{N}$ to the auxiliary space $\{\xi_i,\zeta_i\}_{i=1}^{N}$.
The auxiliary variables are then related to the action-angle variables as
\begin{align}
  \begin{split}
    I_i & = \frac{\xi_i^2 + \zeta_i^2}{2}, \\
    \theta_i & = \tan^{-1}\left(\frac{\xi_i}{\zeta_i}\right) \qquad k = 1, \ldots N-1,
  \end{split}
\end{align}
where $I_i$ is the action variable, and we define $(\xi_N,\zeta_N)$ as the variables of the center of mass.
This transformation is a trivial canonical transformation, as discussed in the previous section.
The entire canonical transformation built by the neural network is illustrated in the main text.
Our canonical transformation consists of (i) point transformations~\cite{Shuo-Hui-etal2020}, (ii) symplectic linear transformations~\cite{Bondesan-Lamacraft2019}, and (iii) the discrete Hartley transformations.
The point transformation is implemented by an invertible neural network. In our approach, we use the RealNVP neural network~\cite{RealNVP2017}.
The scale and the translation transformations in RealNVP are represented by dense neural networks.
We here set the number of hidden layers to $5$ and the dimension of the hidden layer to $30$.
We chose the hyperbolic tangent function as the activation function and stacked $4$ RealNVP.
We implemented the symplectic linear transformation using the method proposed in Ref.~\citen{Bondesan-Lamacraft2019}, in which U(1) transformations were represented by $N$ Householder transformations.
The discrete Hartley transformation is defined by~\cite{Hartley1942,Bracewell1983}
\begin{align}
  \mathrm{DHT}(q)_k \defeq
        \sum_{n=1}^N \sqrt{\frac{1}{N}}q_n\biggl(\cos\left(\frac{2\pi nk}{N}\right) + \sin\left(\frac{2\pi nk}{N}\right)\biggr) \qquad k=1,\ldots,N.\label{eq:discrete_hartley_trans_in_chap4}
\end{align}
The transformation extracts the variables of the center of mass as $\mathrm{DHT}(q)_N$.
For the Toda lattice, the whole canonical transformation was constructed by $10$ elemental blocks consisting of (i) and (ii), and one discrete Hartley transformation layer.

\section{Symplectic Integrator with Adjoint Method}
To calculate the gradients of $L_I$, we used the Yoshida fourth-order symplectic integrator with the adjoint method, which is one of the most popular fourth-order symplectic integrators~\cite{Yoshida1990,Candy-Rozmus1991,Forest-Ruth1990,Neri1987}.
We derived a specific form of the adjoint equations that implement this integrator.
Note that our implementation is a special case of the method proposed in Ref.~\citen{Sanz-Serna2016}.

Let us consider time evolution of a $N$-particle system from $t_{\mathrm{init}}$ to $t_{\mathrm{fin}}$ and $\{t_k\}_{k=1}^{N_{\mathrm{time}}}$ be $N_{\mathrm{time}}$ time points obtained by discretization of the duration.
We also assume that the system Hamiltonian takes the following form:
\begin{align}
    H(p,q) = T(p) + V(q,\theta), \; T(p) = \sum_{i=1}^{N}\frac{p_i^2}{2},
\end{align}
where $V(q,\theta)$ is a learnable potential function, and $\theta$ is determined such that the loss function is minimized.
The equations of motion are
\begin{align}
  \begin{split}
    & \frac{dq_i}{dt} = p_i, \\
    & \frac{dp_i}{dt} = -\frac{\partial V(q,\theta)}{\partial q_i}, \qquad i = 1, \ldots N.
  \end{split}
\end{align}
The Yoshida fourth-order symplectic integrator is represented by~\cite{Yoshida1990}
\begin{align}
  \begin{split}
    & q_i(\tau_{n+1}) = q_i(\tau_{n}) + h c_n p_i(\tau_{n}), \\
    & p_i(\tau_{n+1}) = p_i(\tau_{n}) - h d_n \frac{\partial V(q_i(\tau_{n+1}),\theta)}{\partial q_i}, \qquad n = 1, \ldots 4, \\
  \end{split}\label{eq:Yoshida_algorithm}
\end{align}
where
\begin{align}
  \begin{split}
    & c_1 = c_4 = \frac{1}{2(2-2^{1/3})}, \;   c_2 = c_3 = \frac{1-2^{1/3}}{2(2-2^{1/3})},  \\
    & d_1 = d_3 = \frac{1}{2-2^{1/3}}, \;   d_2 = -\frac{2^{1/3}}{2-2^{1/3}}, \;  d_4 = 0,
  \end{split} \label{eq:coefficients_neri}
\end{align}
$h$ is the time step $(t_{\mathrm{fin}} - t_{\mathrm{init}})/N_\mathrm{time}$, and $\tau_n$ describe intermediate times satisfying $t_k = \tau_1 < \tau_2 < \tau_3 < \tau_4 < \tau_5 = t_{k} + h = t_{k+1}$.
We note that the backward evolution produces exactly the same time series with the forward evolution because of the symmetry of the coefficients.
This property is common for the symplectic integrators constructed by the method proposed in Ref.~\citen{Yoshida1990}.
Let $L$ be the loss function whose minimization determines the parameter $\theta$.
To optimize $L$, we need to calculate the gradients $\lambda_i^q \equiv \frac{\partial L}{\partial q_i}$, $\lambda_i^p \equiv \frac{\partial L}{\partial p_i}$, and $\lambda^\theta \equiv \frac{\partial L}{\partial \theta}$ at $t_{\mathrm{init}}$.
It is known that these gradients satisfy the following adjoint equations~\cite{NeuralODE,Sanz-Serna2016}:
\begin{align}
  \begin{split}
    \frac{d\lambda^{q}_i}{dt} & = \sum_{j=1}^{N}\frac{\partial^2 V(q,\theta)}{\partial q_i q_j} \lambda^{p}_j, \\
    \frac{d\lambda^{p}_i}{dt} & = - \lambda^{q}_i,\\
    \frac{d\lambda^{\theta}}{dt} & =  \sum_{j=1}^{N}\lambda^{p}_j\frac{\partial^2V(q,\theta)}{\partial q_i \partial \theta}.
  \end{split} \label{eq:hamilton_dynamics_of_adjoint}
\end{align}
We obtain the values at $t_{\mathrm{init}}$ by integrating the equation from $t_{\mathrm{fin}}$ to $t_{\mathrm{init}}$ backwards.
As seen in \refeq{eq:hamilton_dynamics_of_adjoint}, the relation between $\lambda^q$ and $\lambda^p$ is equivalent to the relation between the momenta and positions of Hamiltonian mechanics.
Therefore, we can use the symplectic integrator to describe the dynamics of $\lambda^q$ and $\lambda^p$.
Furthermore, the discretization induced by the symplectic integrator produces the equations derived from the chain rule of the algorithm~\cite{Sanz-Serna2016}.
For the Yoshida symplectic integrator, we can see the fact by the following procedure.
The update procedure described by \refeq{eq:Yoshida_algorithm} can be split into two steps:
\begin{align}
  \begin{split}
    q_i(\tau_{n+1}) &  = q_i(\tau_{n}) + h c_n p_i(\tau_{n}), \\
    p_i(\tau_{n}) & = p_i(\tau_{n}), \qquad n = 1, \ldots 4.
  \end{split}
\end{align}
and
\begin{align}
  \begin{split}
    q_i(\tau_{n+1}) & = q_i(\tau_{n+1}), \\
    p_i(\tau_{n+1}) & = p_i(\tau_{n}) - h d_n \frac{\partial V(q_i(\tau_{n+1}),\theta)}{\partial q_i},\qquad n = 1, \ldots 4.
  \end{split}
\end{align}
According to the chain rule, $\lambda^q$ and $\lambda^p$ obey
\begin{align}
  \left(
    \begin{array}{c}
      \lambda^q_{n} \\
      \lambda^p_{n}
    \end{array}
  \right)
  =
  \left(
    \begin{array}{cc}
      I &  0  \\
      hc_n &  I
    \end{array}
  \right)
  \left(
    \begin{array}{c}
      \lambda^q_{n+1} \\
      \lambda^p_{n}
    \end{array}
  \right),\qquad n = 1, \ldots 4,
\end{align}
and
\begin{align}
  \left(
    \begin{array}{c}
      \lambda^q_{n+1} \\
      \lambda^p_{n}
    \end{array}
  \right)
  =
  \left(
    \begin{array}{cc}
      I &  M(\tau_{n+1})  \\
      0 &  I
    \end{array}
  \right)
  \left(
    \begin{array}{c}
      \lambda^q_{n+1} \\
      \lambda^p_{n+1}
    \end{array}
  \right),\qquad n = 1, \ldots 4,
\end{align}
where $M(\tau_{n+1})$ is the matrix whose elements are
\begin{align}
  M_{ij}(\tau_{n+1}) = -hd_n\frac{\partial^2 V(q_i(\tau_{n+1}),\theta)}{\partial q_i q_j}.
\end{align}
The update procedure of $\lambda^q$ and $\lambda^p$ is thus represented by
\begin{align}
  \left(
    \begin{array}{c}
      \lambda^q_{n+1} \\
      \lambda^p_{n+1}
    \end{array}
  \right)
  =
  \left(
    \begin{array}{cc}
      I &  -M(\tau_{n+1})  \\
      0 &  I
    \end{array}
  \right)
  \left(
    \begin{array}{cc}
      I &  0  \\
      -hc_n &  I
    \end{array}
  \right)
  \left(
    \begin{array}{c}
      \lambda^q_{n} \\
      \lambda^p_{n}
    \end{array}
  \right), \qquad n = 1, \ldots 4.
  \label{eq:algorithm_adjoint_symplectic}
\end{align}
This expression is the same as a specific implementation of the symplectic integrator for \refeq{eq:hamilton_dynamics_of_adjoint}.
Since $M(\tau_n)$ depends on $q(\tau_{n})$, we calculate the backward time evolution of $(q,p)$ simultaneously.
The auxiliary variables and $\lambda^{\theta}$ can be regarded as positions and momenta, respectively.
We applied the symplectic integrator to these variables for calculating the time evolution of $\lambda^{\theta}$.

\section{Brief Analysis of Performances}
\begin{figure}[t]
  \centering
  \includegraphics[width=12cm]{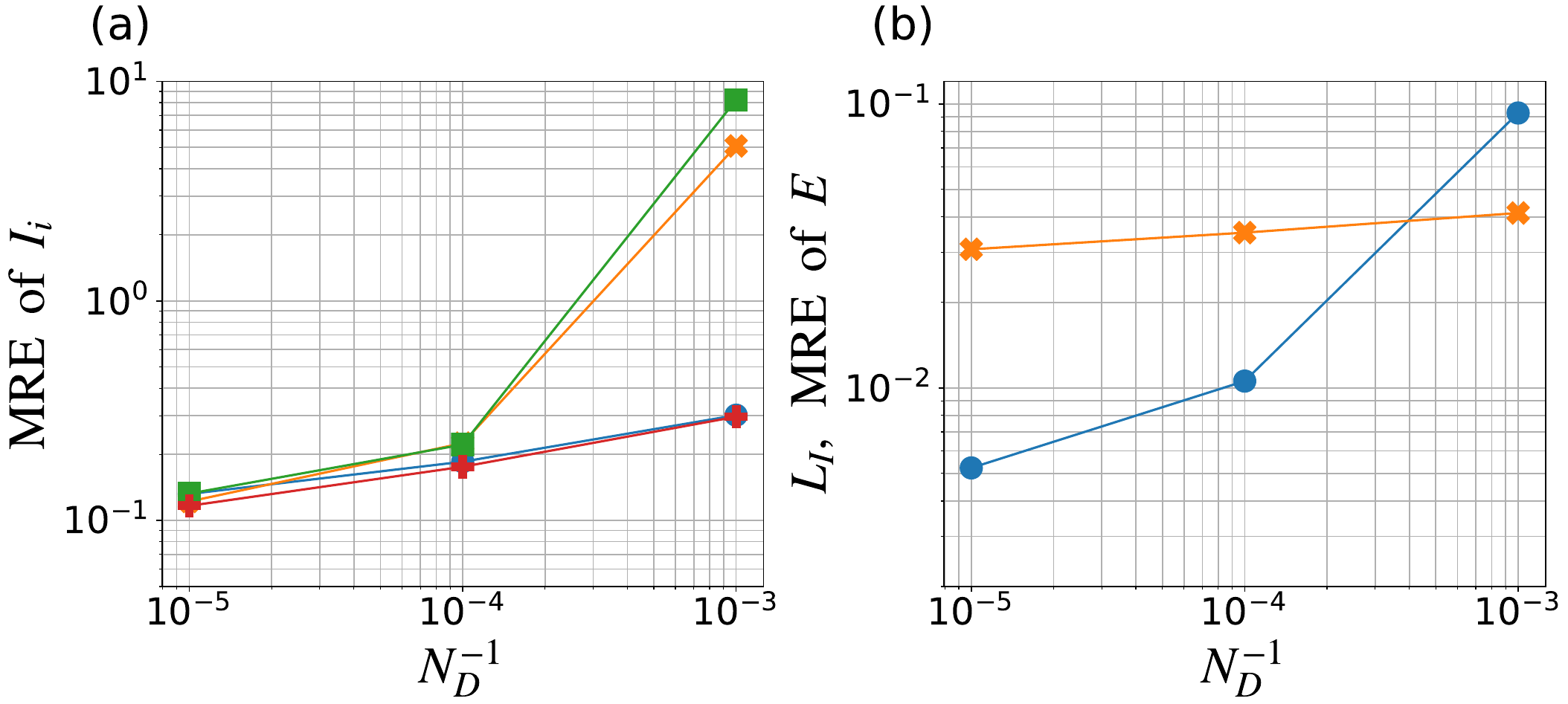}
  \caption{
    (Color online) (a) Training data size dependence of MRE of the action variables: $I_1$ (blue circle), $I_2$ (orange cross), $I_3$ (green square), $I_4$ (red plus) and (b) $L_I$ (blue circle) and MRE of the total energy $E$ (orange cross).
    $N_D$ is the number of the training data.
    $L_I$ is the loss function of the action variables introduced in the Letter.
    }
  \label{fig:toda_larning_residual_result_performance_data_changing}
\end{figure}
\begin{figure}[tp]
  \centering
  \includegraphics[width=12cm]{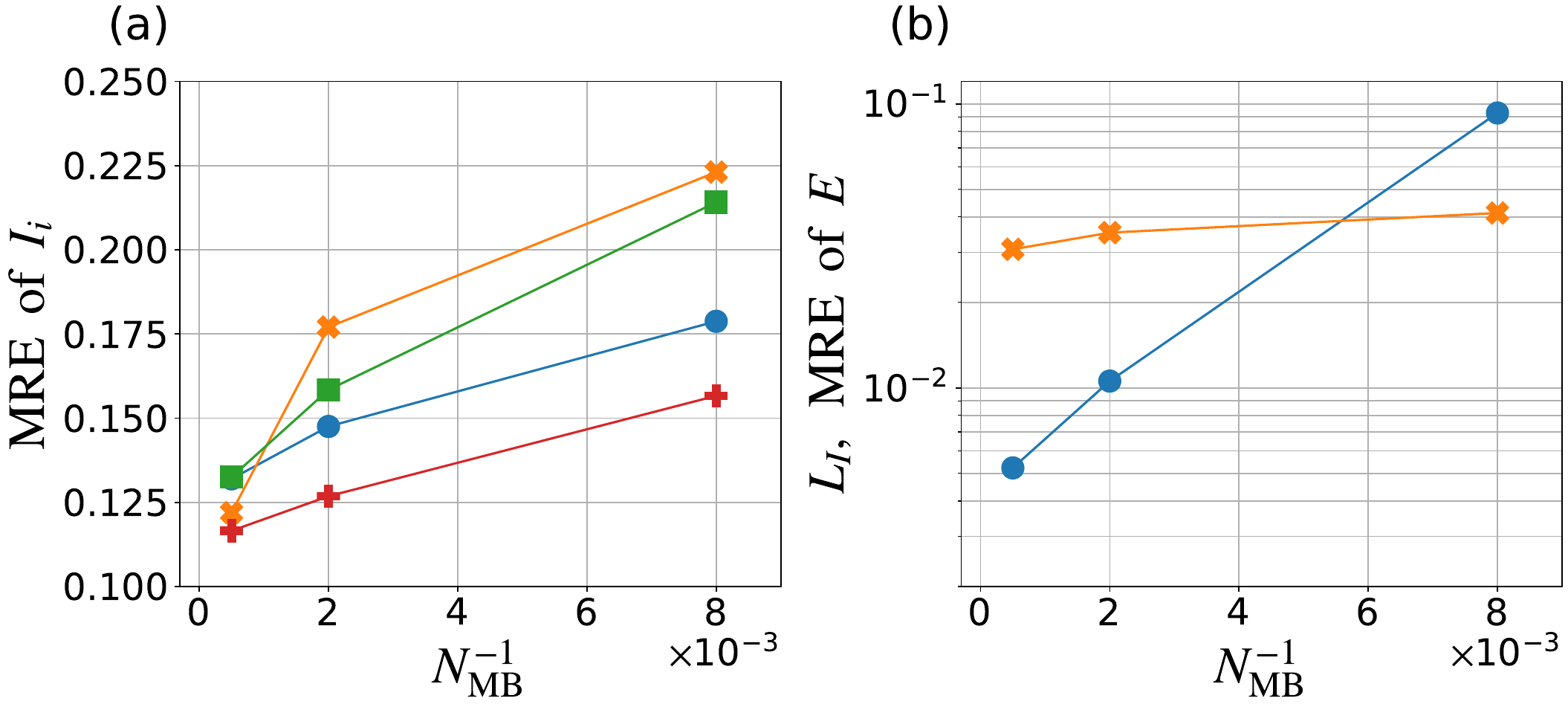}
  \caption{
    (Color online) (a) Mini-batch size dependence of MRE of the action variables: $I_1$ (blue circle), $I_2$ (orange cross), $I_3$ (green square), $I_4$ (red plus) and (b) $L_I$ (blue circle) and MRE of the total energy $E$ (orange cross).
    $N_{\mathrm{MB}}$ stands for the number of the mini-batch.
    $L_I$ is the loss function of the action variables introduced in the Letter.
  }
  \label{fig:toda_larning_residual_result_performance_batch_changing}
\end{figure}

We tested the performance of the trained neural networks by changing the training data size and the mini-batch size.
We set the number of time series for calculations of $L_I$ to $10^3$,
the period of the time development to $100$, the interval for the discretization to $0.01$,
and the number of initial states sampled from the Boltzmann distribution to $10^4$.
The MRE and the loss function for different training data sizes are shown in \reffig{fig:toda_larning_residual_result_performance_data_changing}.
Although the increase of training data monotonically improved the performance, the reduction of the error and the loss became slow for large training data sizes.
This result might be due to the limited representation power of the neural networks.
Nevertheless, the larger the number of parameters and layers in the neural networks is, the smaller the error and the loss should become.
As well as the training data size, we checked the dependence of the performance on the mini-batch size.
The performance for different mini-batch sizes is shown in \reffig{fig:toda_larning_residual_result_performance_batch_changing}.
We observed that the error and the loss were smaller for larger mini-batch sizes.
In the present study, taking these performance analyses into account, we set the training data size to $N_D=10^5$ and the mini-batch size to $N_{MB}=2 \times 10^3$, respectively.

\bibliography{bibliography}